# Order-Optimal Consensus through Randomized Path Averaging

Florence Bénézit*, Alexandros G. Dimakis†,
Patrick Thiran*, Martin Vetterli*†
*School of IC, EPFL, Lausanne CH-1015, Switzerland
†Department of Electrical Engineering and Computer Science (EECS)
University of California, Berkeley
Berkeley, CA 94720, USA

**Abstract**

Gossip algorithms have recently received significant attention, mainly because they constitute simple and robust message-passing schemes for distributed information processing over networks. However for many topologies that are realistic for wireless ad-hoc and sensor networks (like grids and random geometric graphs), the standard nearest-neighbor gossip converges as slowly as flooding ($O(n^2)$ messages).

A recently proposed algorithm called geographic gossip improves gossip efficiency by a $\sqrt{n}$ factor, by exploiting geographic information to enable multi-hop long distance communications. In this paper we prove that a variation of geographic gossip that averages along routed paths, improves efficiency by an additional $\sqrt{n}$ factor and is order optimal ($O(n)$ messages) for grids and random geometric graphs.

We develop a general technique (travel agency method) based on Markov chain mixing time inequalities, which can give bounds on the performance of randomized message-passing algorithms operating over various graph topologies.

## I. Introduction

Gossip algorithms are distributed message-passing schemes designed to disseminate and process information over networks. They have received significant interest because the problem of computing a global function of data distributively over a network, using only localized message-passing, is fundamental for numerous applications.

These problems and their connections to mixing rates of Markov chains have been extensively studied starting with the pioneering work of Tsitsiklis [26]. Earlier work studied mostly deterministic protocols, known as average consensus algorithms, in which each node communicates with each of its neighbors in every round. More recent work (e.g. [12], [2]) has focused on so-called gossip algorithms, a class of randomized algorithms that solve the averaging problem by computing a sequence of randomly selected pairwise averages. Gossip and consensus algorithms have been the focus of renewed interest over the past several years [12], [3], [14], motivated by applications in sensor networks and distributed control systems.

The simplest setup is the following: $n$ nodes are placed on a graph whose edges correspond to reliable communication links. Each node is initially given a scalar (which could correspond to some sensor measurement like temperature) and we are interested in solving the *distributed averaging* problem: namely, to find a distributed message-passing algorithm by which *all nodes* can compute the *average* of all $n$ scalars. A scheme that computes the average can easily be modified to compute any linear function (projection) of the measurements as well as more general functions. Furthermore, the scalars can be replaced with vectors and generalized to address problems like distributed filtering and optimization as well as distributed detection in sensor networks [24], [27], [20]. Random projections computed via gossip, can be used for compressive sensing of sensor measurements and field estimation as proposed in [19]. Note that throughout this paper we will be interested in gossip algorithms that compute linear functions, and will not discuss related problems like information dissemination (see e.g. [15], [21] and references therein).

Gossip algorithms solve the averaging problem by first having each node randomly pick one of their one-hop neighbors and iteratively compute pairwise averages: Initially all the nodes start with their own measurement as an estimate of the average. They update this estimate with a pairwise average of current estimates with a randomly selected neighbor, at each gossip round. An attractive property of gossip is that no coordination is required for the gossip algorithm to converge to the global average when the graph is connected – nodes can just randomly wake up, select one of their one-hop neighbors randomly, exchange estimates and update their estimate with the average. We will refer to this algorithm as *standard* or nearest-neighbor gossip.

A fundamental issue is the performance analysis of such algorithms, namely the communication (number of messages passed between one-hop neighboring nodes) required before a gossip algorithm converges to a sufficiently accurate estimate. For energy-constrained sensor network applications, communication corresponds to energy consumption and therefore should be minimized. Clearly, the convergence time will depend on the graph connectivity, and we expect well-connected graphs to spread information faster and hence to require fewer messages to converge.

This question was first analyzed for the complete graph in [12], where it was shown that $\Theta(n \log \epsilon^{-1})$ gossip messages need to be exchanged to converge to the global average within $\epsilon$ accuracy. Boyd et al. [3] analyzed the convergence time of standard gossip for any graph and showed that it is closely linked to the mixing time of a Markov chain defined on the communication graph. They further addressed the problem of optimizing the neighbor selection probabilities to accelerate convergence.

For certain types of well connected graphs (including expanders and small world graphs), standard gossip converges very quickly, requiring the same number of messages ($\Theta(n \log \epsilon^{-1})$) as the fully connected graph. Note that any algorithm that averages $n$ numbers will require $\Omega(n)$ messages.

Unfortunately, for random geometric graphs and grids, which are the relevant topologies for large wireless ad-hoc and sensor networks, standard gossip is extremely wasteful in terms of communication requirements. For instance, even optimized standard gossip algorithms on grids converge very slowly, requiring $\Theta(n^2 \log \epsilon^{-1})$ messages [3], [8]. Observe that this is of the same order as the energy required for every node to flood its estimate to all other nodes. On the contrary, the obvious solution of averaging numbers on a spanning tree and flooding back the average to all the nodes requires only $O(n)$ messages. Clearly, constructing and maintaining a spanning tree in dynamic and ad-hoc networks introduces significant overhead and complexity, but a quadratic number of messages is a high price to pay for fault tolerance.

Recently Dimakis et al. [8] proposed *geographic gossip*, an alternative gossip scheme that reduces to $\Theta(n^{1.5} \log \epsilon^{-1} / \sqrt{\log n})$ the number of required messages, with slightly more complexity at the nodes. Assuming that the nodes have knowledge of their geographic location and under some assumptions in the network topology, greedy geographic routing can be used to build an *overlay network* where *any pair* of nodes can communicate. The overlay network is a complete graph on which standard gossip converges with $\Theta(n \log \epsilon^{-1})$ iterations. At each iteration we perform greedy routing, which costs $\Theta(\sqrt{n/\log n})$ messages on a geometric random graph. In total, geographic gossip thus requires $\Theta(n^{1.5} \log \epsilon^{-1} / \sqrt{\log n})$ messages.

Li and Dai [13] recently proposed Location-Aided Distributed Averaging (LADA), a scheme that uses partial locations and markov chain lifting to create fast gossiping algorithms. The cluster-based LADA algorithm performs slightly better than geographic gossip, requiring $\Theta(n^{1.5} \log \epsilon^{-1} / (\log n)^{1.5})$ messages for random geometric graphs. While the theoretical machinery is different, LADA algorithms also use directionality to accelerate gossip, but can operate even with partial location information and have smaller total delay compared to geographic gossip, at the cost of a somewhat more complicated algorithm.

*This paper:* We investigate the performance of *path averaging*, which is the same algorithm as geographic gossip with the additional modification of *averaging all the nodes on the routed paths*. Observe that averaging the whole route comes almost for free in multihop communication, because a packet can accumulate the sum and the number of nodes visited, compute the average when it reaches its final destination and follow the same route backwards to disseminate the average to all the nodes along this route.

In path averaging, the selection of the routed path (and hence the routing algorithm) will affect the performance of the algorithm. We start this paper by experimentally observing that the number of messages for grids and random geometric graphs seems to scale linearly when random greedy routing is used.

The mathematical analysis of path averaging with greedy routing is highly complex because the number of possible routes increases exponentially in the number of nodes. To make the analysis tractable we make two simplifications: a) We eliminate edge effects by assuming a grid or random geometric graph on a torus b) we use *box-greedy routing*, a scheme very similar to greedy routing with the extra restriction that each hop is guaranteed to be within a virtual box that is not too close or too far from the existing node. Box-greedy routing (described in section III-D) can be implemented in a distributed way if each node knows its location, the location of its one-hope neighbors, and the total number of nodes $n$. We call path averaging with box-greedy routing *Box-path averaging*.

The main result of this paper is that geographic gossip with path averaging requires $O(n)$ messages under these assumptions. Further, we present experimental evidence that suggests that this optimal behavior is preserved even when different routing algorithms are used.

The remainder of this paper is organized as follows: in Section II we define our time and network models, give a precise definition of gossip algorithms and explain our metrics to evaluate the performance of gossip algorithms.

| | Grid | Random geometric graph |
|---|---|---|
| Standard gossip [3] | $\mathcal{C}_{ave} = \Theta(n^2 \log \epsilon^{-1})$ | $\mathcal{C}_{ave} = \Theta\left(\frac{n^2 \log \epsilon^{-1}}{\log n}\right)$ |
| Hops per time-slot | $\mathbb{E}[R] = \Theta(\sqrt{n})$ | $\mathbb{E}[R] = \Theta\left(\sqrt{\frac{n}{\log n}}\right)$ |
| Geographic | $T_{ave} = \Theta(n \log \epsilon^{-1})$ | $T_{ave} = \Theta(n \log \epsilon^{-1})$ |
| gossip [8] | $\mathcal{C}_{ave} = \Theta(n^{1.5} \log \epsilon^{-1})$ | $\mathcal{C}_{ave} = \Theta\left(\frac{n^{1.5} \log \epsilon^{-1}}{\sqrt{\log n}}\right)$ |
| Box- | $T_{ave} = \Theta(\sqrt{n} \log \epsilon^{-1})$ | $T_{ave} = \Theta(\sqrt{n \log n} \log \epsilon^{-1})$ |
| path averaging | $\mathcal{C}_{ave} = \Theta(n \log \epsilon^{-1})$ | $\mathcal{C}_{ave} = \Theta(n \log \epsilon^{-1})$ |

TABLE I

PERFORMANCE OF DIFFERENT GOSSIP ALGORITHMS. $T_{ave}$ DENOTES $\epsilon$-AVERAGING TIME (IN GOSSIP ROUNDS) AND $\mathcal{C}_{ave}$ DENOTES EXPECTED NUMBER OF MESSAGES REQUIRED TO ESTIMATE WITHIN $\epsilon$ ACCURACY.

In Section III, we describe path averaging with greedy routing and show its excellent performance in simulations. We also define path averaging with box-greedy routing (box-path averaging), whose analysis is tractable and gives insight on general gossip algorithms. In Section IV we present the technical tools we use to theoretically show the efficiency of box-path averaging. We show that the methodology developed in that section is general, simple and insightful. Section IV-D states our results, and outlines the proofs which can be found in the Appendix.

## II. BACKGROUND AND METRICS

### A. Time model

We use the asynchronous time model [1], [3], which is well-matched to the distributed nature of sensor networks. In particular, we assume that each sensor has an independent clock whose "ticks" are distributed as a rate $\lambda$ Poisson process. However, our analysis is based on measuring time in terms of the number of ticks of an equivalent single virtual global clock ticking according to a rate $n\lambda$ Poisson process. An exact analysis of the time model can be found in [3]. We will refer to the time between two consecutive clock ticks as one timeslot.

Throughout this paper we will be interested in minimizing the number of messages without worrying about delay. We can therefore adjust the length of the timeslots relative to the communication time so that only one packet exists in the network at each timeslot with high probability. Note that this assumption is made only for analytical convenience; in a practical implementation, several packets might co-exist in the network, but the associated congestion control issues are beyond the scope of this work.

### B. Network model

We model the wireless networks as random geometric graphs (RGG), following standard modeling assumptions [11], [18]. A random geometric graph $G(n, r)$ is formed by choosing $n$ node locations uniformly and independently in the unit square, with any pair of nodes $i$ and $j$ connected if their Euclidean distance is smaller than some transmission radius $r$ (see Fig. 1). It is well known [18], [11], [10] that in order to maintain connectivity and to minimize interference, the transmission radius $r(n)$ should scale like $r(n) = \sqrt{c \log n / n}$. For the purposes of analysis, we assume that communication within this transmission radius always succeeds[1]. Note that we assume that the messages involve real numbers; the effects of message quantization in gossip and consensus algorithms, is an active area of research (see for example [17], [25]).

In the Appendix we show a slightly stronger condition than connectivity, on how the scaling coefficient $c$ in $r(n)$ tunes the regularity of random geometric graphs. The result states that, if $c > 10$, then a random geometric graph is *regular* with high probability when $n$ is large. Regular geometric graphs are random geometric graphs with degrees bounded above and below. In particular, select constants $a < \alpha < b$, draw a random geometric graph and divide the unit square in squares of size $\alpha \log n / n$. If each square contains between $a \log n$ and $b \log n$ nodes, then the graph is called regular. One standard result [11], [18] that for a suitable constant $\alpha$, each of these squares will contain one or more nodes with high probability (w.h.p.). In the appendix we prove a slightly stronger regularity condition: that in fact, if $\alpha > 2$, the number of nodes in each square will be $\Theta(\log n)$ nodes, i.e. the random geometric graphs are regular geometric graphs w.h.p. In Section III-D, we assume that our network is a regular geometric graph embedded on a torus, and we ensure that any node in a square is able to communicate with any other node of its four neighboring squares by setting $c > 10$.

### C. Gossip algorithms

Gossip is a class of distributed averaging algorithms, where average consensus can be reached up to any desired level of accuracy by iteratively averaging small random groups of estimates. At time-slot $t = 0, 1, 2, \ldots$, each node $i = 1, \ldots, n$ has an estimate $x_i(t)$ of the global average. We use $x(t)$ to denote the $n$-vector of these estimates and therefore $x(0)$ gathers the initial values to be averaged. The ultimate goal is to drive the estimate $x(t)$ to the vector of averages $\bar{x}_{\text{ave}}\vec{1}$, where $\bar{x}_{\text{ave}} := \frac{1}{n}\sum_{i=1}^{n} x_i(0)$, and $\vec{1}$ is an $n$-vector of ones. In gossip, at each time-slot $t$, a random set $S(t)$ of nodes communicate with each other and update their estimates to the average of the estimates of $S(t)$: for all $j \in S(t)$, $x_j(t+1) = \sum_{i \in S(t)} x_i(t)/|S(t)|$. In standard gossip (nearest neighbor) and in geographic gossip, only random pairs of nodes average their estimates, hence $S(t)$ always contains exactly two nodes. On the other hand, in path averaging, $S(t)$ is the set of nodes in the random route generated at each time-slot $t$. Therefore in this case, $S(t)$ contains a random number of nodes.

[1] However, we note that our proposed algorithm remains robust to communication and node failures.

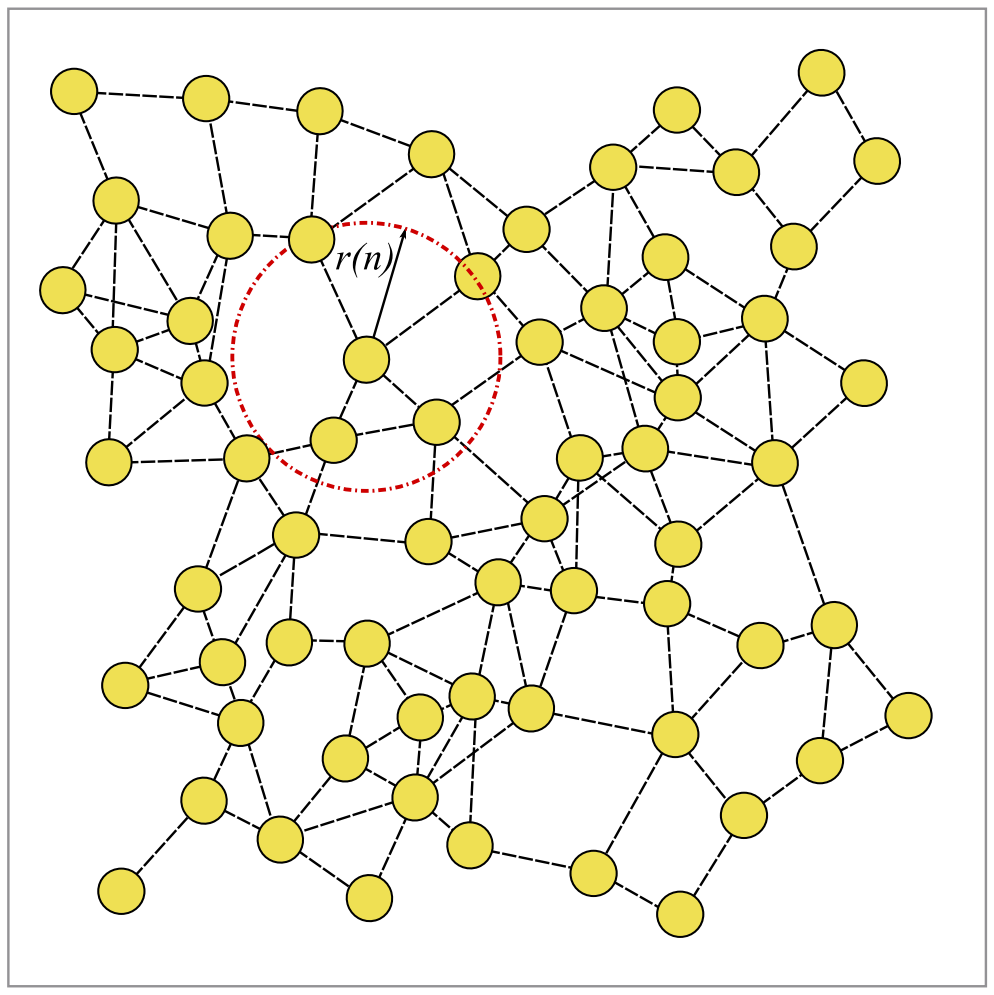

Fig. 1. Random geometric graph example. The connectivity radius is $r(n)$.

### D. Metrics for convergence time and message cost

We measure the performance of gossip algorithms with a metric that was recently introduced in [6]. Instead of defining convergence time as the time $T_{ave}$ elapsed until the error metric becomes smaller than $\epsilon$ with probability $1-\epsilon$ (see Eq. (2)) as in [3], we define it as the time $T_c$ by which the error metric is divided by an $e$ factor with probability $1$ in the long run. Apart from giving an almost sure criterion for convergence time, consensus time $T_c$ also conveniently lightens the formalism by removing the $\epsilon$'s.

For the algorithms of interest, the estimate vector $x(t)$ and the error vector $\varepsilon(t) = x(t) - \bar{x}_{\text{ave}}\vec{1}$ for $t > 0$ are *random*. However, in the long run, the error decays exponentially with a *deterministic* rate $1/T_c$, where $T_c$, called consensus time, is defined as follows [6]:

*Theorem 1: Consensus time $T_c$.* If $\{S(t)\}_{t\geqslant 0}$ is an independently and identically distributed (i.i.d.) process, then the limit

$$-\frac{1}{T_c} = \lim_{t\to\infty} \frac{1}{t} \log \|\varepsilon(t)\|, \tag{1}$$

where $\|\cdot\|$ denotes the $\ell_2$ norm, exists and is a constant with probability $1$.

In other words, after a transient regime, the number of iterations needed to reduce the error $\|\varepsilon\|$ by a factor $e$ is almost surely equal to $T_c$, which therefore characterizes the speed of convergence of the algorithm. $T_c$ is easy to measure in experiments, and can be theoretically upper bounded. However lower bounding this quantity remains an open problem.

Previous work defined the $\epsilon$-averaging time $T_{ave}(\epsilon)$, another quantity describing speed of convergence [3] (see also [9] for a related analysis):

*Definition 1: $\epsilon$-averaging time $T_{ave}(\epsilon)$.* Given $\epsilon > 0$, the $\epsilon$-averaging time is the earliest time at which the vector

$x(k)$ is $\epsilon$ close to the normalized true average with probability greater than $1-\epsilon$:

$$T_{ave}(\epsilon) = \sup_{x(0)} \inf_{t=0,1,2\ldots} \left\{ \mathbb{P}\left( \frac{\|x(t) - x_{ave}\vec{1}\|}{\|x(0)\|} \geq \epsilon \right) \leq \epsilon \right\}. \quad (2)$$

Although $T_{ave}(\epsilon)$ is hard to measure in practice because it requires the evaluation of an infinite number of probabilities, it is easily upper and lower bounded theoretically in terms of the spectral gap (see Section IV). Indeed $T_{ave}(\epsilon)$ contains a probability tolerance $\epsilon$ in its definition, which facilitates greatly its analysis. On the contrary, $T_c$ is hard to analyze theoretically because it is constrained by the exigency of its inherent determinism. An important issue is the behavior of $T_c$ and $T_{ave}$ as the number $n$ of nodes in the network grows. It can be shown that $T_c(n) = O(T_{ave}(n,\epsilon))$ for any fixed $\epsilon$, but whether the two quantities are equivalent and under which conditions is still an open problem. Previous theoretical results summarized in Table I refer to $\epsilon$-averaging time.

We compare algorithms in terms of the amount of required communication. More specifically, let $R(t)$ represent the number of one-hop radio transmissions required in time-slot $t$. In a standard gossip protocol, the quantity $R(t) \equiv R$ is simply a constant, whereas for our protocol, $\{R(t)\}_{t\geqslant 1}$ will be a sequence of i.i.d. random variables. The total communication cost at time-slot $t$, measured in one-hop transmissions, is given by the random variable $C(t) = \sum_{k=1}^{t} R(k)$. Consensus cost $\mathcal{C}_c$ is defined as follows [6]:

*Theorem 2: Consensus cost* $\mathcal{C}_c$. If $\{S(t)\}_{t\geqslant 0}$ is an independently and identically distributed (i.i.d.) process, then the following limit exists and is a constant with probability 1:

$$\begin{aligned} -\frac{1}{\mathcal{C}_c} &= \lim_{t\to\infty} \frac{1}{C(t)} \log \|\varepsilon(t)\| \\ &= \lim_{t\to\infty} \frac{t}{C(t)} \lim_{t\to\infty} \frac{\log \|\varepsilon(t)\|}{t}. \end{aligned}$$

Thus, $\mathcal{C}_c = \mathbb{E}[R(1)]T_c$ is the number of one-hop transmissions needed in the long run to reduce the error by a factor $e$ with probability 1.

Similarly, we define the expected $\epsilon$-averaging cost $\mathcal{C}_{ave}(\epsilon)$ to be the *expected* communication cost in the first $T_{ave}(\epsilon)$ iterations of the algorithm: $\mathcal{C}_{ave}(\epsilon) = \mathbb{E}[C(T_{ave}(\epsilon))] = \mathbb{E}[R(1)]T_{ave}(\epsilon)$.

## III. Path averaging algorithms

### A. Path averaging on random geometric graphs.

The proposed algorithm combines gossip with random greedy geographic routing. A key assumption is that each node knows its location and is able to learn the geographic locations of its one-hop neighbors (for example using a single transmission per node). Also the nodes need to know the size of the space they are embedded in. Note that while our results are developped for random geometric topologies, the algorithm can be applied on any set of nodes embedded on some compact and convex region.

The algorithm operates as follows: at each time-slot one random node activates and selects a random position (target) on the unit square region where the nodes are spread out. Note that no node needs to be located on the target, since this would require global knowledge of locations. The node then creates a packet that contains its current estimate of the average, its position, the number of visited nodes so far (one), the target location, and passes the packet to a neighbor that is *randomly chosen among its neighbors closer to the target*. As nodes receive the packet, randomly and greedily forwarding it towards the target, they add their value to the sum and increase the hop counter. When the packet reaches its destination node (the first node whose nearest neighbors have larger distance to the target compared to it), the destination node computes the average of all the nodes on the path, and reroutes that information backwards on the same route. See Fig. 2 for an illustration of random greedy routing. It is not hard to show [8] that for $G(n, r)$ when $r$ scales like $\Theta(\sqrt{\log n/n})$, greedy forwarding succeeds to reach the closest node to the random target with high probability over graphs — in other words there are no large 'holes' in the network. We will refer to this whole procedure of routing a message and averaging on a random path as one gossip round which lasts for one time-slot, after which $O(\sqrt{n/\log n})$ nodes will replace their estimates with their joint average. We prefer not to route the estimates by choosing the next node as the *closest* neighbor to the target, but as one random neighbor *closer* to the target, because we observed that the latter is cheaper (smaller $C_c$). Note that the nodes do not need to know the number of nodes $n$ in the network, they only need the size of the field on which they are deployed.

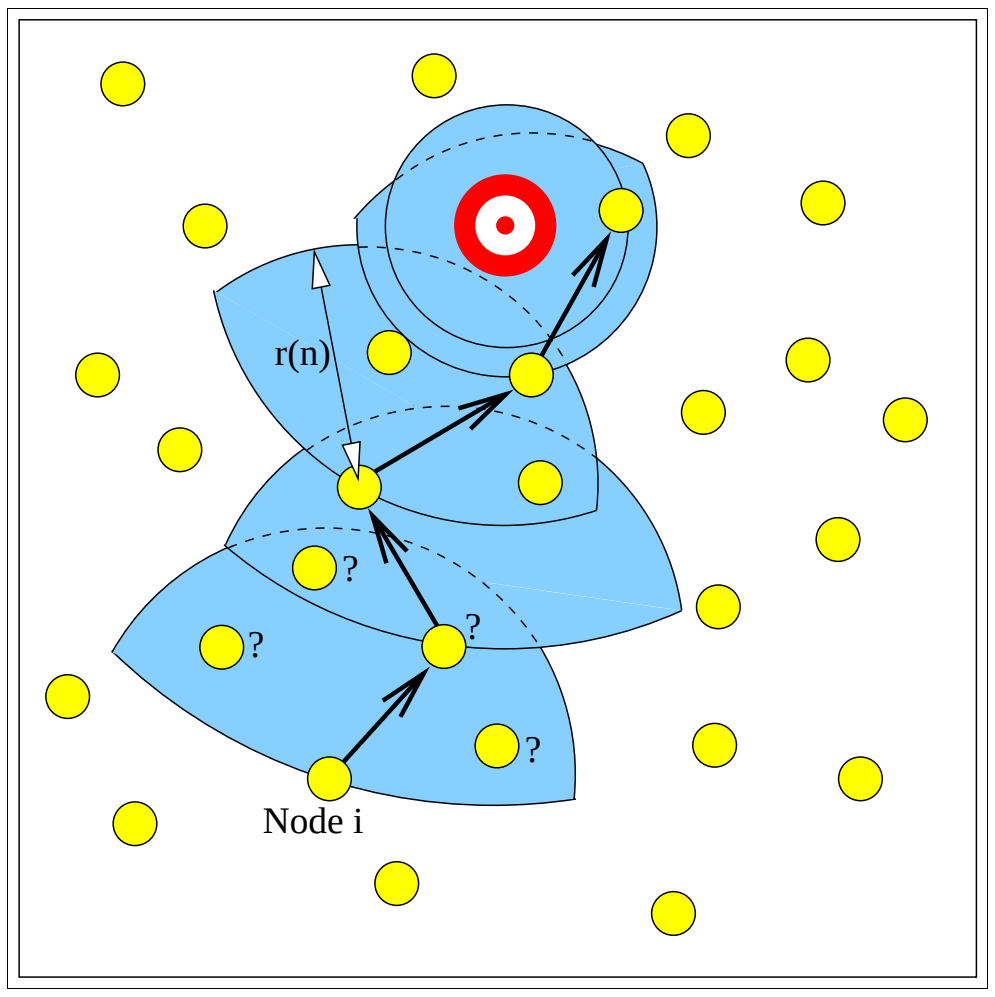

Fig. 2. Random greedy routing. Node $i$ has to choose the following node in the route among the nodes that are his neighbors (inside the ball of radius $r(n)$ centered in node $i$) *and* that are closer to the target than $i$ (inside the ball of radius centered in the target, where $d$ is the distance between node $i$ and the target). Next node is thus randomly chosen in the intersection of the two balls.

## *B. Motivation–Performance simulations*

We experimentally measured $T_c$ and $\mathcal{C}_c$ in order to evaluate the performance of path averaging on random geometric graphs with a growing number $n$ of nodes in the unit square. Fig 3(b) shows that our algorithm behaves strikingly better than standard gossip and geographic gossip, when, for example, $r(n) = \sqrt{c \log n/n}$ with $c = 4.5$. For other values of $c$, the performance of our algorithm also greatly improves previous gossip schemes. Most importantly, for small connection radius $r(n)$ (small $c$), the number of messages $\mathcal{C}_c$ behaves almost linearly in $n$

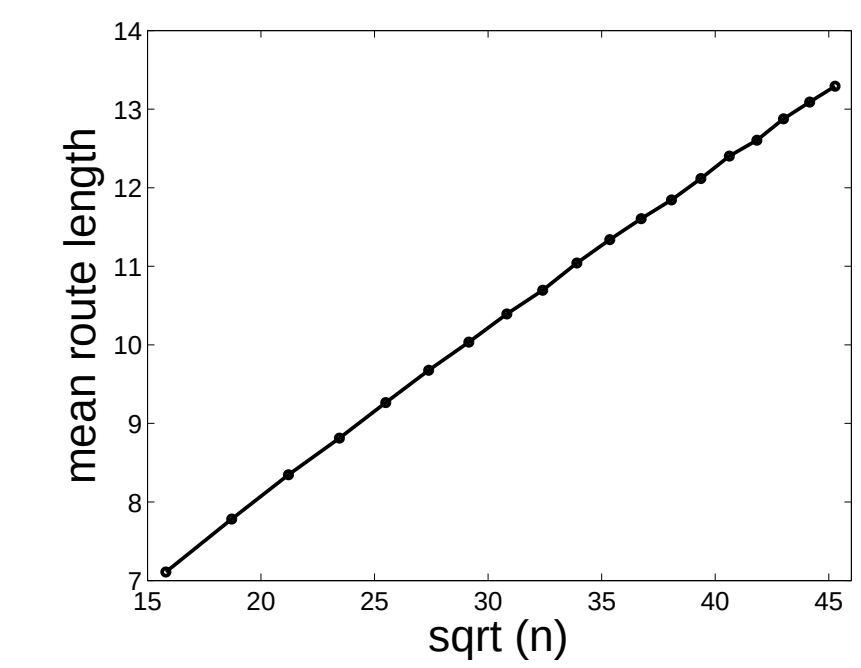

(a) Mean route length $\mathbb{E}(R)$.

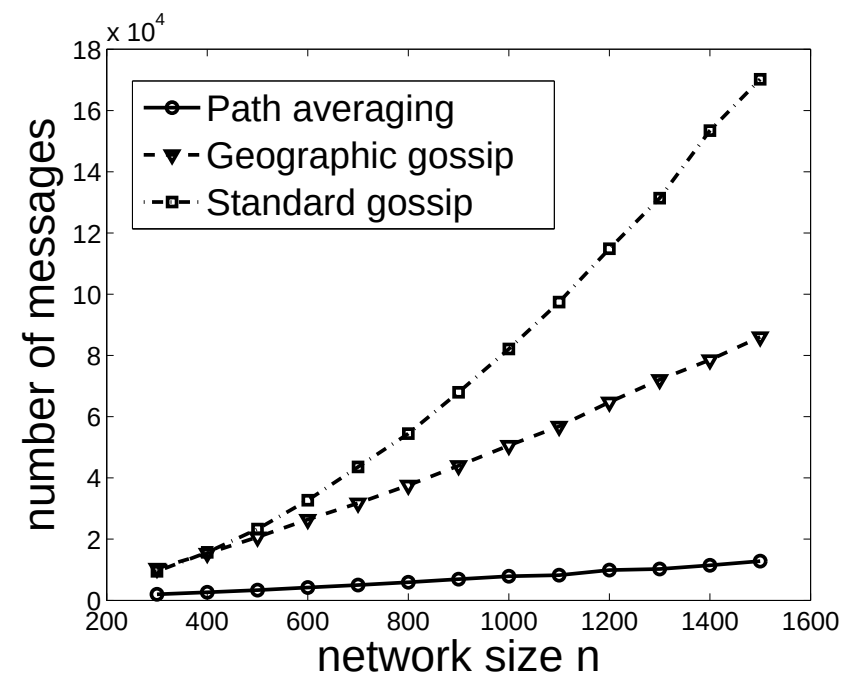

(b) Consensus cost $\mathcal{C}_c$: compare three methods

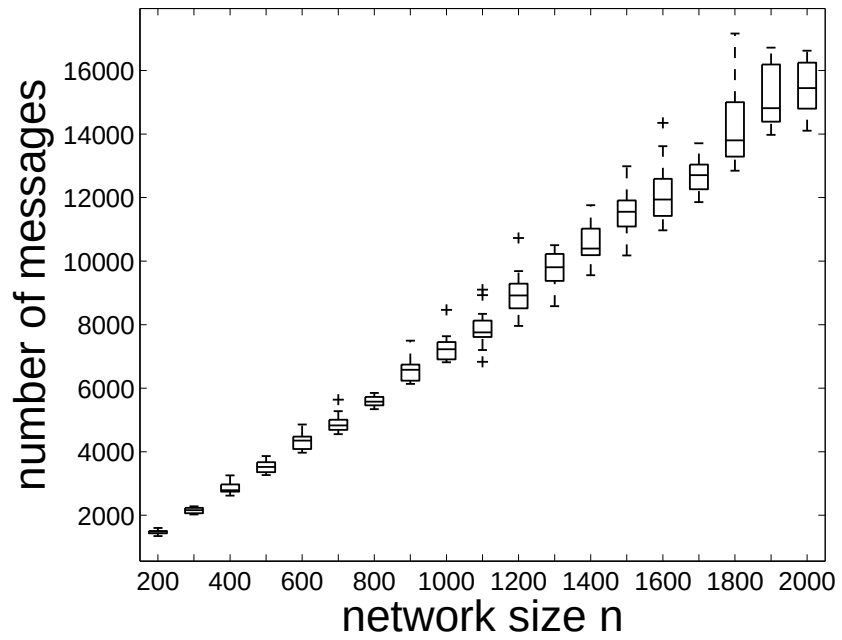

(c) $\mathcal{C}_c$: path averaging, $r(n) = \sqrt{4.5 \log n/n}$

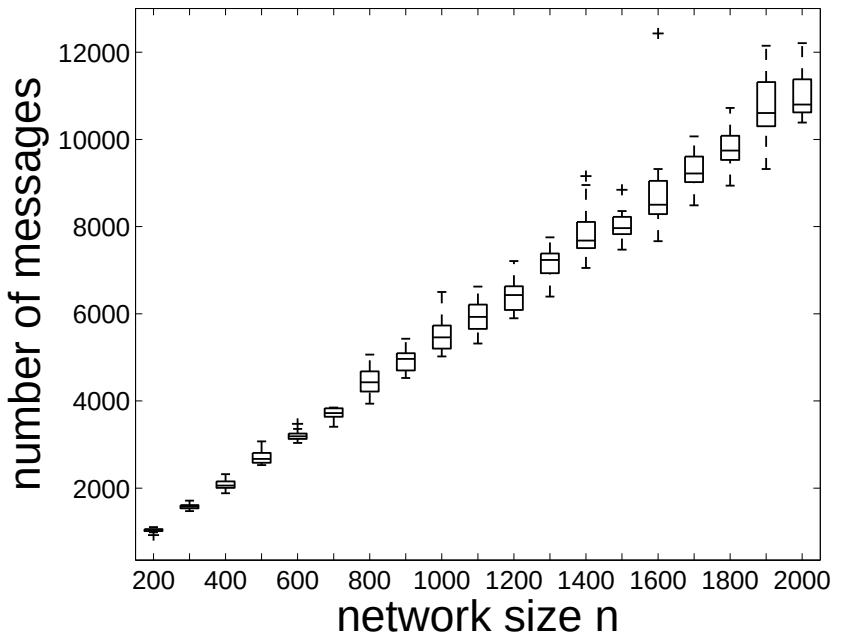

(d) $\mathcal{C}_c$: path averaging, $r(n) = \sqrt{25 \log n/n}$

Fig. 3. Performance of path averaging. The simulations were performed over 15 graphs per $n$. Averaging time was measured here by $T_c \simeq (t_1 - t_2)/[\log \|\epsilon(t_2)\| - \log \|\epsilon(t_1)\|]$ for $t_1 = 500$ and $t_2 = 1750$. (a) The mean route length in random greedy routing behaves in $\sqrt{n/\log n}$. (b) Comparison between standard gossip, geographic gossip (without rejection sampling) and path averaging with $r(n) = \sqrt{4.5 \log n/n}$. (c), (d) Consensus costs $\mathcal{C}_c = \mathbb{E}[R]T_c$ for radii $r(n) = \sqrt{4.5 \log n/n}$ and $r(n) = \sqrt{25 \log n/n}$.

(see Fig. 3(c)), and as $c$ increases, the behavior improves (see Fig. 3(d)). The slight super-linearity in Fig.3(c) is due to small $r(n)$ and possibly edge effects. Clearly, we cannot expect better than linear behavior in $n$ because at least $n$ messages are necessary to average $n$ values. Therefore path averaging with greedy routing seems to be optimal for sufficiently large constant $c$.

Unfortunately, the theoretical analysis of path averaging with greedy routing seems intractable. However, with a slight modification in the routing algorithm, and by ignoring edge effects, we are able to analyze path averaging, first for grids and then for regular geometric graphs. Recall that random geometric graphs are regular geometric graphs with high probability when $n$ large if $c$ is sufficiently large (Section II-B).

### *C. $(\leftrightarrow, \updownarrow)$-path averaging on grids*

The first step in our analysis is understanding the behavior of path averaging on regular grids using a simple routing scheme. Throughout this paper, a grid of $n$ nodes will be a 4-connected lattice on a torus of size $\sqrt{n} \times \sqrt{n}$. $(\leftrightarrow, \updownarrow)$-path averaging performs as follows: At each iteration $t$, a randomly selected node $I$ wakes up and selects

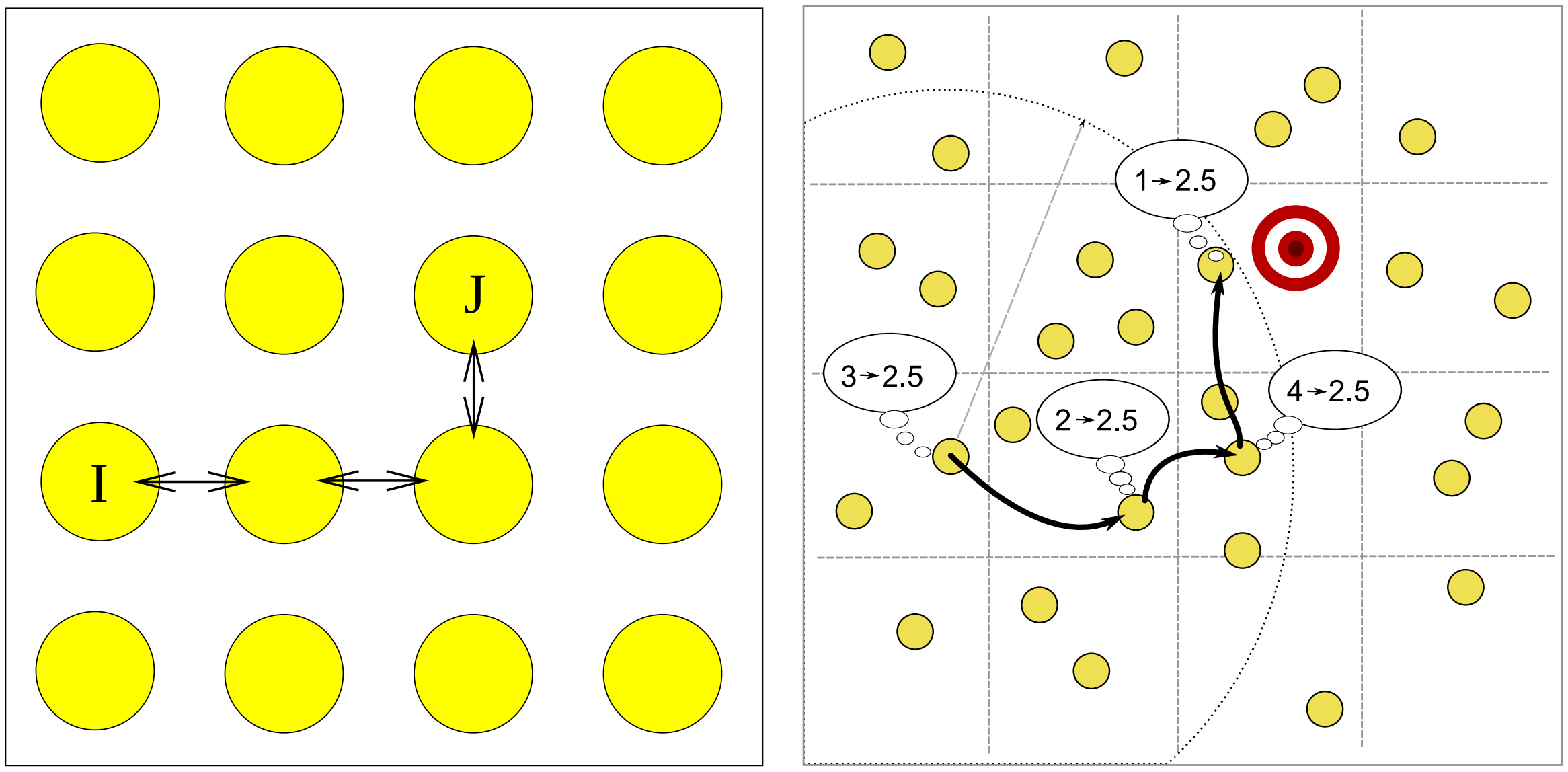

(a) $(\leftrightarrow, \updownarrow)$-route (b) box-path averaging

Fig. 4. (a) Shortest $(\leftrightarrow, \updownarrow)$-route from $I$ to $J$ on the grid. (b) Example of box-path averaging on an RGG: The node with inital value 3 selects a random position and places a target. Using $(\leftrightarrow, \updownarrow)$-box routing towards that target, all the nodes on the path replace their values with the average of the four nodes.

a random destination node $J$ so that the pair $(I, J)$ is independently and uniformly distributed. Node $I$ also flips a fair coin to design the first direction: horizontal $(\leftrightarrow)$ or vertical $(\updownarrow)$. If for instance horizontal was picked as the first direction, the path between $I$ and $J$ is then defined by the shortest horizontal-vertical route between $I$ and $J$ (see Fig. 4(a)). The estimates of all the nodes on this path are aggregated and averaged by messages passed on this path, and at the end of the iteration the estimates of the nodes on this path are updated to their global average. Clearly, this message-passing procedure can be executed if each node knows its location on the grid.

### *D. Box-path averaging on regular geometric graphs*

As seen in Section II-B, a regular geometric graph can be organized in virtual squares with the transmission radius $r(n)$ selected so that a node can pass messages to any node in the four squares adjacent to its own square.

In box-path averaging, when a node activates, it chooses uniformly at random a target location in the unit torus and its initial direction: horizontal or vertical. Then a node is selected uniformly from the ones in the adjacent square in the right direction. (Recall that regularity ensures that w.h.p. $\Theta(\log n)$ nodes will be in each square.) The routing stops when the message reaches a node in the square where the target is located. As in the previous path averaging algorithms, the estimates of all the nodes on the path are averaged and all the nodes replace their values with this estimate (see Fig. 4(b)). The key point is that box-path averaging can be executed if each node knows its location, the locations of its one-hop neighbors and the total number of nodes $n$, because with this knowledge each node can figure out which square it belongs to and pass messages appropriately.

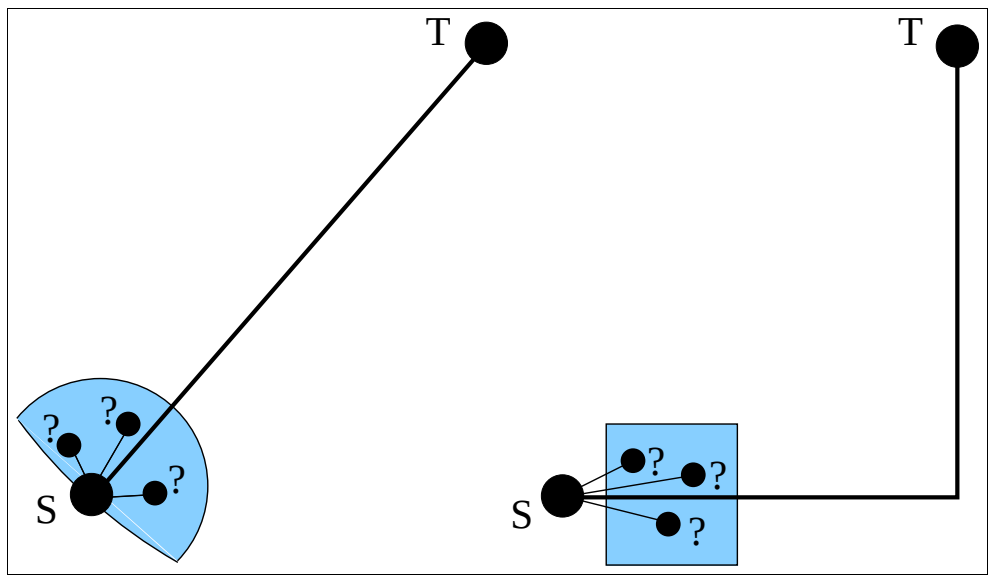

Fig. 5. Choosing next node in the route. On the left: random greedy routing, on the right: ($\updownarrow, \leftrightarrow$)-box routing. It is easy to see that the two choice areas contain on average $\Theta(\log n)$ nodes.

Box-greedy routing is a regularized version of random greedy routing, and is introduced to make the analysis tractable. Both routing schemes proceed by choosing the next hop among $\Theta(\log n)$ nodes (Fig. 5). Box-greedy routing generates routes with $\Theta(\sqrt{n/\log n})$ hops on average, and random greedy routing does as well on experiments (Fig. 3(a)). We are now ready to start the theoretical analysis of the aforementioned path averaging algorithms.

## IV. Analysis

### A. *Averaging and eigenvalues.*

Let $x(t)$ denote the vector of estimates of the global averages after the $t^{th}$ gossip round, where $x(0)$ is the vector of initial measurements. Any gossip algorithm can be described by an equation of the form

$$x(t+1) = W(t)x(t), \tag{3}$$

where $W(t)$ is the averaging matrix over the $t^{th}$ time-slot.

We say that the algorithm converges almost surely (a.s.) if $P[\lim_{t\to\infty} x(t) = x_{ave}\vec{1}] = 1$. It converges in expectation if $\lim_{t\to\infty} \mathbb{E}[x(t) - x_{ave}\vec{1}] = 0$, and there is mean square convergence if $\lim_{t\to\infty} \mathbb{E}[\|x(t) - x_{ave}\vec{1}\|_2] = 0$. There are two *necessary* conditions for convergence:

$$\begin{cases} \vec{1}^T W(t) = \vec{1}^T \\ W(t)\vec{1} = \vec{1}, \end{cases} \tag{4}$$

which respectively ensure that the average is preserved at every iteration, and that $\vec{1}$ is a fixed point. For any linear distributed averaging algorithm following (3) where $\{W(t)\}_{t\geq 0}$ is i.i.d., conditions for convergence in expectation and in mean square can be found in [2]. In gossip algorithms, $W(t)$ are symmetric and projection matrices. Taking into account this particularity, we can state specific conditions for convergence. Let $\lambda_2(\mathbb{E}[W])$ be the second largest eigenvalue in magnitude of the expectation of the averaging matrix $\mathbb{E}[W] = \mathbb{E}[W(t)]$. If condition (4) holds and if $\lambda_2(\mathbb{E}[W]) < 1$, then $x(t)$ converges to $x_{ave}\vec{1}$ in expectation and in mean square.

In the case where $\{W(t)\}_{t\geq 0}$ is stationary and ergodic (and thus in particular when $\{W(t)\}_{t\geq 0}$ is i.i.d.), sufficient conditions for a.s. convergence can be proven [5]: if the gossip communication network is connected, then the estimates of gossip converge to the global average $\bar{x}_{\text{ave}}$ with probability 1. More precisely, define $T_\eta := \inf\{t \geq 1 : \prod_{p=0}^{t} W(t-p) \geq \eta > 0\}$. $T_\eta$ is a stopping time. If $\mathbb{E}[T_\eta] < \infty$, then the estimates converge to the global average with probability 1. In other words, every node has to eventually connect to the network, which has to be jointly connected.

Interestingly, the value of $\lambda_2(\mathbb{E}[W])$, that appears in the criteria of convergence in expectation and of mean square convergence, controls the speed of convergence:

$$T_c(\mathbb{E}[W]) \leqslant \frac{2}{\log\left(\frac{1}{\lambda_2(\mathbb{E}[W])}\right)} \leq \frac{2}{1-\lambda_2(\mathbb{E}[W])}. \tag{5}$$

A straightforward extension of the proof of Boyd et al. [3] from the case of pairwise averaging matrices to the case of symmetric projection averaging matrices yields the following bound on the $\epsilon$-averaging time, which also involves $\lambda_2(\mathbb{E}[W])$:

$$T_{ave}(\epsilon, \mathbb{E}[W]) \leq \frac{3\log\epsilon^{-1}}{\log\left(\frac{1}{\lambda_2(\mathbb{E}[W])}\right)} \leq \frac{3\log\epsilon^{-1}}{1-\lambda_2(\mathbb{E}[W])}. \tag{6}$$

There is also a lower bound of the same order, which implies that $T_{ave}(\epsilon, \mathbb{E}[W]) = \Theta(\log\epsilon^{-1}/(1-\lambda_2(\mathbb{E}[W])))$.

Consequently, the rate at which the *spectral gap* $1-\lambda_2(\mathbb{E}[W])$ approaches zero as $n$ increases, controls both the $\epsilon$-averaging time $T_{ave}$ and the consensus time $T_c$. For example, in the case of a complete graph and uniform pairwise gossiping, one can show that $\lambda_2(\mathbb{E}[W]) = 1 - 1/n$. Therefore, as previously mentioned, the consensus time of this scheme is $O(n)$. In pairwise gossiping, the convergence time and the number of messages have the same order because there is a constant number $R$ of transmissions per time-slot. In geographic gossip and in path averaging on random geometric graphs, one round uses many messages for the path routing ($\sqrt{n/\log n}$ messages on average), hence multiplying the order of consensus time $T_c(n)$ by $\sqrt{n/\log n}$ gives the order of consensus cost $\mathcal{C}_c(n)$.

## B. The travel agency method

A direct consequence of the previous section is that the evaluation of consensus time requires an accurate upper bound on $\lambda_2(\mathbb{E}[W])$. Consequently, computing the averaging time of a scheme takes two steps: (1) evaluation of $\mathbb{E}[W]$, (2) upperbound of its second largest eigenvalue in magnitude. $\mathbb{E}[W]$ is a doubly stochastic matrix that corresponds to a time-reversible Markov Chain.

We can therefore use techniques developed for bounding the spectral gap of Markov Chains to bound the convergence time of gossip. In particular, we will use Poincaré's inequality by Diaconis and Stroock [7] (see also [4], p. 212-213 and the related canonical paths technique [23]) to develop a bounding technique for gossip.

*Theorem 3 (Poincaré's inequality [7]):* Let $P$ denote an $n \times n$ irreducible and reversible stochastic matrix, and $\pi$ its left eigenvector associated to the eigenvalue 1 ($\pi^T P = \pi^T$) such that $\sum_{i=1}^{n} \pi(i) = 1$. A pair $e = (k, l)$ is called an edge if $P_{kl} \neq 0$. For each ordered pair $(i, j)$ where $1 \leqslant i, j \leqslant n$, $i \neq j$, choose one and only one path $\gamma_{ij} = (i, i_1, \ldots, i_m, j)$ between $i$ and $j$ such that $(i, i_1)$, $(i_1, i_2)$, ..., $(i_m, j)$ are all edges. Define

$$|\gamma_{ij}| = \frac{1}{\pi(i)P_{ii_1}} + \frac{1}{\pi(i_1)P_{i_1i_2}} + \ldots + \frac{1}{\pi(i_m)P_{i_mj}}. \tag{7}$$

The Poincaré coefficient is defined as

$$\kappa = \max_{\text{edge } e} \sum_{\gamma_{ij} \ni e} |\gamma_{ij}| \pi(i)\pi(j). \tag{8}$$

Then the second largest eigenvalue of $P$ verifies

$$\lambda_2(P) \leq 1 - \frac{1}{\kappa}. \tag{9}$$

We will apply this theorem with $P = \mathbb{E}[W]$. Here $\pi(i) = 1/n$ for all $1 \leqslant i \leqslant n$.

The combination of Poincaré inequality with bounds 5 and 6 forms a versatile technique for bounding the performance of gossip algorithms that we call the *travel agency* method. It is crucial to understand that the edges used in the application of the theorem are abstract and do not correspond to actual edges in the physical network. They instead correspond to paths on which there is joint averaging, and hence information flow, through message-passing. Consider the following analogy. Imagine that $n$ airports are positioned at the locations of the nodes of the network. In this scenario, we are given a table $P = \mathbb{E}[W]$ of the flight capacities (number of passengers per time unit) between any pair of airports among the $n$ airports. A good *averaging intensity* $\mathbb{E}[W_{ij}]$ between nodes $i$ and $j$ correspond to a good *capacity* flight between airports $i$ and $j$ in the travel agency method. Here edges $e$ are existing flights and, in our specific case, there is the same number of travelers in all the airports ($\pi(i) = 1/n$ for all $i$). We are asked to design one and only one road map $\gamma_{ij}$ between each pair of airports $i$ and $j$ that avoids congestion and multiple hops. $|\gamma_{ij}|$ measures the level of congestion between airport $i$ and airport $j$. The theorem tells us that if we can come up with a road map that avoids significant congestion on the worst flight (i.e. if $\kappa$ is small), then we will have proven that the flying network is efficient ($\lambda_2$ is small). The previous bounds 5,6 can now be used to bound the consensus time and consensus cost.

One of the important benefits of this bounding technique is that we do not need know the entries of $\mathbb{E}[W]$ to bound the averaging cost, and only good lower bounds suffice. In terms of the analogy, we only need to know that each flight $(i, j)$ has at least capacity $C_{i,j}$. If $(i, j)$ can actually carry more passengers ($P_{i,j} \geqslant C_{i,j}$), then our measure of congestion $\kappa$ will be overestimated. While our final upper-bounds will not be as tight as they could have been if we had exact knowledge of $\mathbb{E}[W]$, they suffice to establish the optimal asymptotic behavior.

### C. *Example: standard gossip revisited*

In order to illustrate the generality of our technique, we show how to apply it on simple examples, by giving sketches of novel proofs for known results on nearest neighbors gossip on the complete graph and on the random geometric graph.

*1) Complete graph:* For any $i \neq j$, $\mathbb{E}[W_{ij}] = 1/n^2$. Indeed $W_{ij} = 0.5$ when node $i$ wakes up (event of probability $1/n$) and chooses node $j$ (event of probability $1/n$ as well), or when $j$ wakes up and chooses $i$. We apply now the travel agency method. We see in $\mathbb{E}[W]$ that all flights have equal capacity $1/n^2$ and that there are direct flights between any pair of airports. We choose here the simplest road map one could think of: to go from airport $i$ to airport $j$, each traveller should take the direct hop $\gamma_{ij} = (i,j)$. Then the sum in (7) has only one term: $|\gamma_{ij}| = n^3$. In this case all flights are equal and one flight $e = (i,j)$ belongs only to one road map: $\gamma_{ij}$. Thus the sum in (8) also has only one term and $\kappa = n^3/(n \cdot n) = n$. Therefore $\lambda_2(\mathbb{E}[W]) \leqslant 1 - 1/n$, which proves that $T_c(n) = O(n)$. Note that the complete graph is the overlay network of geographic gossip [2] (every pair of node can be averaged at the expense of routing), which thus performs in $\mathcal{C}_c(n) = O(n\sqrt{n/\log n})$.

*2) Random geometric graph (RGG):* We show in the Appendix that if the connection radius $r(n)$ is large enough, then RGGs are regular with high probability, i.e. the nodes are very regularly spread out in the unit square, which implies that each node has $\Theta(\log n)$ neighbors. To keep the illustration of the travel agency method simple, we assume that the nodes lie on a torus (no border effects). Consider the pair of nodes $(i,j)$. If $i$ and $j$ are not neighbors, then $\mathbb{E}[W_{ij}] = 0$; if $i$ and $j$ are neighbors, then $\mathbb{E}[W_{ij}] = \Theta\left(1/(n \log n)\right)$ because node $i$ wakes up with probability $1/n$ and chooses node $j$ with probability $\Theta\left(1/\log n\right)$. We now have to create a roadmap with only short distance paths. Regularity ensures that there are no isolated nodes that could create local congestion. We thus naturally decide that the best way to go is to select paths along the straightest possible line between the departure airport and the destination airport. This will require $O(\sqrt{n/\log n})$ hops, therefore the right hand side of Equation (7) is the sum of $O(\sqrt{n/\log n})$ terms, each of equal order:

$$|\gamma_{ij}| = O\left(\sqrt{\frac{n}{\log n}}\right)\frac{1}{1/n}\Theta\left(\frac{1}{1/n\log n}\right) = O(n^2\sqrt{n\log n}). \tag{10}$$

Now we need to compute in how many paths each particular flight is used. It follows from our regularity and torus assumptions that each flight appears in approximatively the same number of road maps. There are $n^2$ paths that use $O(\sqrt{n/\log n})$ flights, but there are only $\Theta(n \log n)$ different flights, hence each flight is used in $O\left((n/\log n)^{1.5}\right)$ paths. We can now compute the Poincaré coefficient $\kappa$. We drop the $\max_e$ argument in Equation (8) because all flights are equal. As $\pi(i) = \pi(j) = 1/n$,

$$\begin{aligned} \kappa &= \sum_{\gamma_{ij}\ni e} O(n^2\sqrt{n\log n})\frac{1}{n}\frac{1}{n} && (11)\\ &= O\left((\frac{n}{\log n})^{1.5}\right) O(\sqrt{n\log n}) && (12)\\ &= O(\frac{n^2}{\log n}), && (13) \end{aligned}$$

which proves that $T_c(n) = O(n^2/\log n)$.

*3) Comments:* The proof of the performance of path averaging on a RGG given in Section B gives insight on how to complete this last proof. It is interesting to see that the travel agency method describes how information will *diffuse* in the network. In the second example, far away nodes will never directly average their estimates, but they will do it indirectly, using the nodes between them.

Note that our method does not give lower-bounds on $\lambda_2(\mathbb{E}[W])$, which would be useful to give an equivalent order for $\epsilon$-averaging time $T_{ave}$. In the case of path averaging, this is not an issue since it is not possible to achieve better than the consensus cost $\mathcal{C}_c(n) = \Theta(n)$. So if the method shows that $T_c(n) = O(\sqrt{n \log n})$, we have that $\mathcal{C}_c(n) = O(\sqrt{n \log n}) O(\sqrt{n / \log n}) = O(n)$ and we can conclude that $\mathcal{C}_c(n) = \Theta(n)$.

[2]In reality, geographic gossip will not be completely uniform but rejection sampling can be used [8] to tamper the distribution

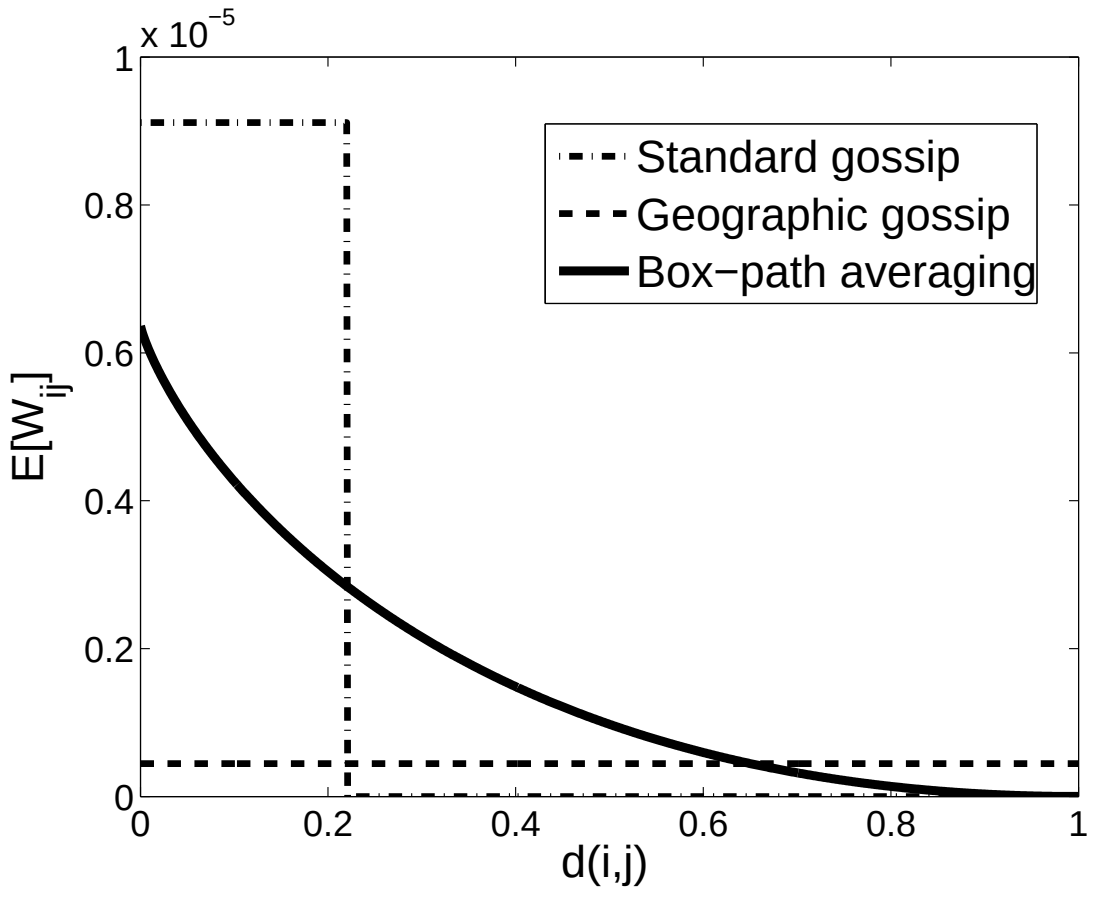

Fig. 6. Behavior of $\mathbb{E}[W_{ij}]$ as a function of the distance in norm 1 between $i$ and $j$ for standard gossip, geographic gossip and box-path averaging.

### *D. Main Results*

The main results of this paper is that the consensus cost of $(\leftrightarrow, \updownarrow)$-path averaging on grids and of box-path averaging on random geometric graphs, behave *linearly* in the number of nodes $n$:

*Theorem 4 ($(\leftrightarrow, \updownarrow)$-path averaging on grids):* On a $\sqrt{n} \times \sqrt{n}$ torus grid, the consensus time $T_c(n)$ of $(\leftrightarrow, \updownarrow)$-path averaging, described in Section III-C, is $O(\sqrt{n})$. Furthermore, the consensus cost is linear: $\mathcal{C}_c(n) = O(n)$.

*Theorem 5 (Box-path averaging on RGG):* Consider a random geometric graph $G(n, r)$ on the unit torus with $r(n) = \sqrt{\frac{c \log n}{n}}$, $c > 10$. With high probability over graphs, the consensus time $T_c(n)$ of box-path averaging, described in Section III-D, is $O(\sqrt{n \log n})$. Furthermore, the consensus cost is linear: $\mathcal{C}_c(n) = O(n)$.

The proofs of Theorem 4 and Theorem 5 are given in the Appendix. Both proofs have the same structure: we first lower bound the entries of $\mathbb{E}[W]$ and next upper bound its second largest eigenvalue in magnitude. Figure 6 shows the behavior of $\mathbb{E}[W_{ij}]$ as a function of the $L_1$ distance between nodes $i$ and $j$ for standard gossip, geographic gossip and path averaging; respectively the proofs give us the insight behind the good performance of box-path averaging compared to standard gossip and geographic gossip by simply analysing Fig. 6. Box-path averaging concentrates the *averaging intensities* $\mathbb{E}[W_{ij}]$ of node $i$ in the area of nodes $j$ close to $i$. Indeed, the closer two nodes, the higher the probability that they are on the same route. Thus, as we can observe on Fig. 6, close nodes have a much higher averaging intensity $\mathbb{E}[W_{ij}]$ than in geographic gossip, where nodes are equally rarely averaged together (the proof shows an order $\sqrt{n / \log n}$ higher). However, the averaging intensity gained by close nodes is lost for far away nodes, which do not average together well anymore (a factor $n$ loss compared to geographic gossip).

In terms of the travel agency method, in box-path averaging over the unit area torus, flights with that cover distances shorter than $1/2$ have high capacity, whereas long distance flights are rare. To apply the method, the idea is to chose 2-hop paths: to go from node $i$ to node $j$, the path will contain two hops that stop half way, in order to exclusively and fairly use the high capacity flights. Remember that standard gossip needs $\sqrt{n/\log n}$ flights per path (see Section IV-C.1), which heavily penalizes the performance despite a very high averaging intensity $\mathbb{E}[W_{ij}]$ for neighboring nodes $i$ and $j$ (see Fig. 6, where $\mathbb{E}[W_{ij}]$ is large for neighboring nodes but falls to $0$ for distances larger than $r(n)$). The performance of path averaging algorithms is good thanks to a diffusion scheme requiring only $O(1)$ flights in each path and $O(1)$ uses of each flight in the road map, combined with a high enough level of averaging intensity $\mathbb{E}[W_{ij}]$. Each node can act as a diffusion relay for some far away nodes, so that the whole network can benefit from the concentration of the averaging intensity.

As a summary, in contrast with geographic gossip, path averaging and standard gossip *concentrate* their averaging intensity on close nodes, which leads to larger coefficients $\mathbb{E}[W_{i,j}]$ when nodes $i$ and $j$ are close enough. However, while standard gossip pays for its concentration with long paths overusing every existing flight, the diffusion pattern of path averaging operates in $2$ steps only without creating any congestion (more precisely, we compute in the proof that each flight is used in at most $9$ paths). In conclusion, the analysis shows that path averaging achieves a good tradeoff between promoting *local* averaging to increase averaging intensity (large $\mathbb{E}[W_{ij}]$) and favoring *long distance* averaging to get an efficient diffusion pattern (every path $\gamma_{ij}$ contains only $O(1)$ edges, and every edge $e$ appears in only $O(1)$ paths).

## V. Conclusions

We introduced a novel gossip algorithm for distributed averaging. The proposed algorithm operates in a distributed and asynchronous manner on locally connected graphs and requires an order-optimal number of communicated messages for random geometric graph and grid topologies. The execution of path averaging requires that each node knows its own location, the locations of its nearest-hop neighbors and (for the routing-scheme that was theoretically analyzed) the total number of nodes $n$.

Location information is independently useful and likely to exist in many application scenarios. The key idea that makes path averaging so efficient is the opportunistic combination of routing and averaging. The issues of delay (how several paths can be concurrently averaged in the network) and fault tolerance (robustness and recovery in failures) remain as interesting future work.

More generally, we believe that the idea of greedily routing towards a randomly pre-selected target (and processing information on the routed paths) is a very useful primitive for designing message-passing algorithms on networks that have some geometry. The reason is that the target introduces some directionality in the scheduling of message passing which avoids diffusive behavior. Other than computing linear functions, such path-processing algorithms can be designed for information dissemination or more general message passing computations such as marginal computations or MAP estimates for probabilistic graphical models [22]. Scheduling the message-passing using some form of linear paths can accelerate the communication required for the convergence of such algorithms. We plan to investigate such protocols in future work.

## VI. Definitions

### A. Notation

- $G(n, r)$ or $RGG$: random geometric graph with $n$ nodes and connection radius $r$.
- $x(0)$: vector of the initial values to be averaged.
- $\bar{x}_{\text{ave}} = \sum_{k=1}^{n} x_k(0)/n$.
- $x(t)$: vector of the estimates of the average.
- $S(t)$: the random set of nodes that average together at time-slot $t$.
- $R(t)$: number of one hop transmissions at time-slot $t$.
- $\epsilon(t) = x(t) - \bar{x}_{\text{ave}}\vec{1}$: error vector, where $\vec{1}$ is the vector of all ones.
- $W(t)$: averaging matrix at time $t$.
- $\lambda_2$: second largest eigenvalue in magnitude.
- $\gamma_{ij}$: path starting in $i$ and ending in $j$.
- $|\gamma_{ij}|$ measures the "resistance" of path $\gamma_{ij}$ (Eq. (7)).
- $\kappa$: Poincaré coefficient (Eq. (8)).
- $T_{ave}(\epsilon)$: $\epsilon$-averaging time (Def. 2)
- $C_{ave}(\epsilon) = \mathbb{E}[R(1)]T_{ave}$: expected $\epsilon$-averaging cost.
- $T_c$, $C_c$: consensus time, consensus cost (Def. 1, 2).

### B. List of the algorithms

- Standard gossip: pairwise gossip where only direct neighbors can average their estimates together.
- Geographic gossip: pairwise gossip where any pair of nodes can average their estimates together at the expense of routing.
- Path averaging: at each iteration a random route is created by random greedy routing in an RGG. The nodes of the route average their estimates together.
- $(\leftrightarrow, \updownarrow)$-path averaging: at each iteration a random route is created by $(\leftrightarrow, \updownarrow)$-routing on a grid (embedded on a torus in the analysis). The nodes of the route average their estimates together.
- Box-path routing: at each iteration a random route is created by box-routing on a regular geometric graph (embedded on a torus in the analysis). The nodes of the route average their estimates together.

# Appendix

## A. Performance of $(\leftrightarrow,\updownarrow)$-path averaging on a grid

This section prooves Theorem 4, which states the linearity of consensus cost for $(\leftrightarrow,\updownarrow)$-path averaging on a grid. The analyzed algorithm is described in Section III-C.

*1) Notation:* We need to define the shortest distance on a torus. To this end, we introduce a torus absolute value $|.|_{\mathcal{T}}$ and a torus $L_1$ norm $\|.\|_1$. For any algebraic value $x$ on a one dimensional torus (circle with $\sqrt{n}$ nodes) and any vector $i$ on a $\sqrt{n}\times\sqrt{n}$ two dimensional torus,

$$\begin{aligned} |x|_{\mathcal{T}} &= \min(|x|, |x-\sqrt{n}|, |x+\sqrt{n}|) \\ \|i\|_1 &= |i_x|_{\mathcal{T}} + |i_y|_{\mathcal{T}}. \end{aligned}$$

We call $\ell_{ij} = \|j-i\|_1$ the $L_1$ distance between nodes $i$ and $j$. The shortest routes between $I$ and $J$ have $\alpha = \ell_{IJ}+1 = |J_x - I_x|_{\mathcal{T}} + |J_y - I_y|_{\mathcal{T}} + 1$ nodes to be averaged, thus the non-zero coefficients of their corresponding matrices $W$ are all equal to $1/\alpha$.

To each route $r$, we assign a generalized gossip $n\times n$ matrix $W^{(r)}$ that averages the current estimates of the nodes on the route. Consequently, at iteration $t$, $W(t) = W^{(r(t))}$, where $r(t)$ was randomly chosen. We call $R$ the route random variable, $s(R)$ its starting node, $d(R)$ its destination node, and $\ell(R) = \ell_{s(R)d(R)}+1$ its number of nodes. As we choose the shortest route, the maximum number of nodes a route can contain is $\sqrt{n}$ if $\sqrt{n}$ is odd, $\sqrt{n}+1$ if $\sqrt{n}$ is even, which can be written as $2\lfloor\sqrt{n}/2\rfloor+1$ in short.

*2) Evaluating $\mathbb{E}[W]$:*

*Lemma 1: (Expected $\mathbb{E}[W]$ on the grid)* For any pair of nodes $(i,j)$, if their distance normalized to the maximum distance $\delta_{ij} = \|j-i\|_1/\sqrt{n}$ is smaller than a constant, then

$$\mathbb{E}[W_{i,j}] = \Omega\left(\frac{1}{n^{1.5}}\right). \tag{14}$$

More precisely,

$$\mathbb{E}[W_{i,j}] \geq \frac{2(1-\delta_{ij}+\delta_{ij}\log\delta_{ij})}{n\sqrt{n}}.$$

Therefore, as expected, far away nodes are less likely to be jointly averaged compared to neighboring ones (see Figure 6). *Proof:* Observing that $\mathbb{E}[W^{(R)}|(\leftrightarrow,\updownarrow)] = \mathbb{E}[W^{(R)}|(\updownarrow,\leftrightarrow)]$ because the route from a node $I$ to a node $J$ horizontally first has the same nodes as the route from $J$ to $I$ vertically first, we get

$$\begin{aligned} \mathbb{E}[W] &= \mathbb{E}[W^{(R)}] \\ &= \frac{1}{2}\mathbb{E}[W^{(R)}|(\leftrightarrow,\updownarrow)] + \frac{1}{2}\mathbb{E}[W^{(R)}|(\updownarrow,\leftrightarrow)] \\ &= \mathbb{E}[W^{(R)}|(\leftrightarrow,\updownarrow)]. \end{aligned}$$

So, for a given pair of nodes $(i,j)$, we can compute the $(i,j)$th entry of the matrix expectation $\mathbb{E}[W]$ by systematically routing first horizontally. Only the $(\leftrightarrow,\updownarrow)$-routes which contain both these two nodes $i$ and $j$ will have a non-zero contribution in $\mathbb{E}[W_{ij}]$. Pick such a route $r$, the $(i, j)$th entry of the corresponding averaging matrix is $W^{(r)}_{i,j} = 1/\ell(r)$. We call $\mathcal{R}^{\ell}_{ij}$ the set of $(\leftrightarrow, \updownarrow)$-routes with $\ell$ nodes passing by node $i$ and by node $j$, and denote $x^+ = \max(x, 0)$. It is not hard to see that $(\ell - \ell_{ij})^+$ is the number of routes of length $\ell$ passing by $i$ first and $j$ next (see Fig. 7), so $|\mathcal{R}^{\ell}_{ij}| = 2(\ell - \ell_{ij})^+$. We thus have for any $i \neq j$:

$$\begin{aligned}
\mathbb{E}[W_{i,j}] &= \sum_r W^{(r)}_{i,j} \mathbb{P}[R = r] \\
&= \frac{1}{n^2} \sum_r W^{(r)}_{i,j} \\
&= \frac{1}{n^2} \sum_{\ell=\ell_{ij}+1}^{2\lfloor \frac{\sqrt{n}}{2} \rfloor + 1} \frac{|\mathcal{R}^{\ell}_{ij}|}{\ell} \\
&= \frac{2}{n^2} \sum_{\ell=\ell_{ij}+1}^{2\lfloor \frac{\sqrt{n}}{2} \rfloor + 1} \frac{\ell - \ell_{ij}}{\ell},
\end{aligned}$$

from which we can deduce that for $i \neq j$

$$\begin{aligned}
\mathbb{E}[W_{i,j}] &\leq \frac{2}{n^2} \int_{\ell_{ij}+1}^{\sqrt{n}+2} \frac{x - \ell_{ij}}{x} dx \\
&= \frac{2}{n^2} \left( \sqrt{n} - \ell_{ij} + 1 - \ell_{ij} \ln \frac{\sqrt{n}+2}{\ell_{ij}+1} \right) \\
\mathbb{E}[W_{i,j}] &\geq \frac{2}{n^2} \int_{\ell_{ij}}^{\sqrt{n}} \frac{x - \ell_{ij}}{x} dx \\
&= \frac{2}{n^2} \left( \sqrt{n} - \ell_{ij} - \ell_{ij} \ln \frac{\sqrt{n}}{\ell_{ij}} \right).
\end{aligned}$$

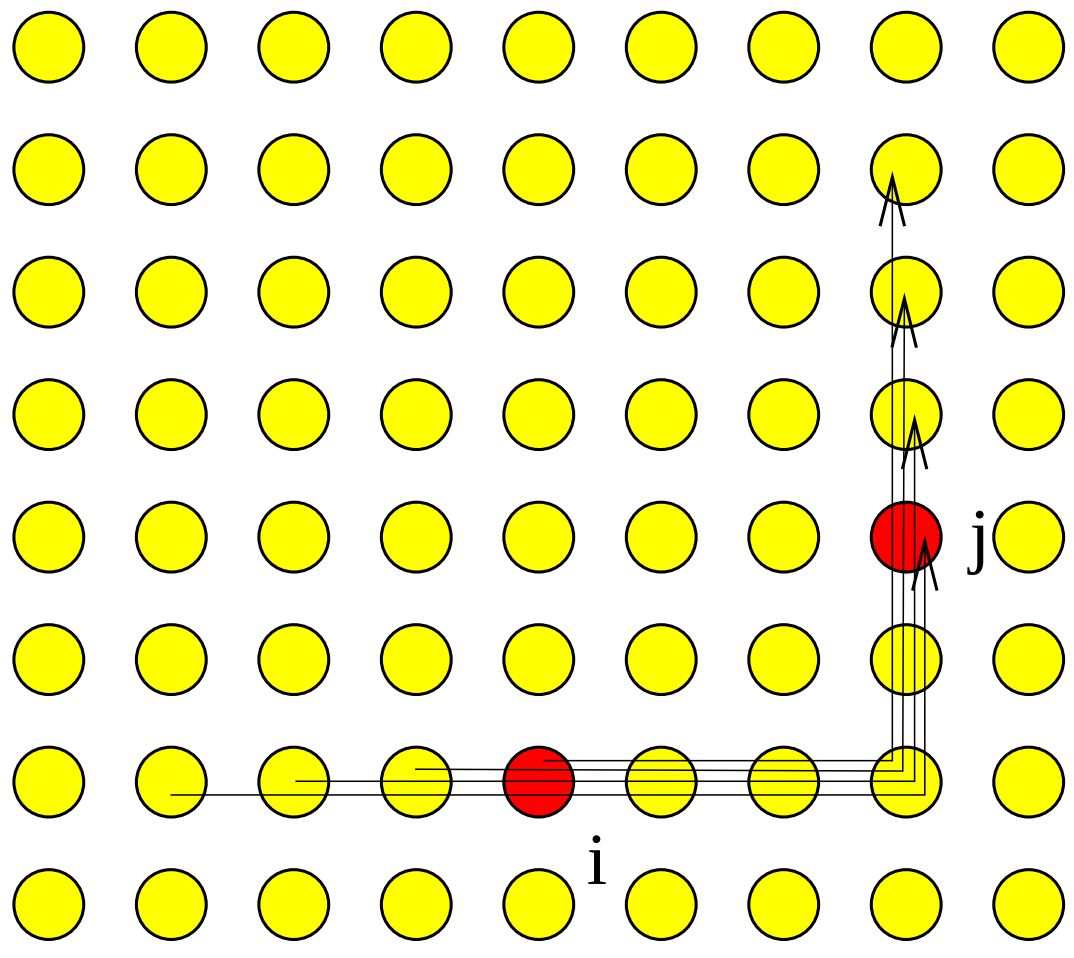

Fig. 7. Counting the number of routes of length $\ell = 9$ nodes, in the case where $\ell_{ij} = 5$. There are $\ell - \ell_{ij} = 9 - 5 = 4$ possible routes with exactly $\ell$ nodes going through node $i$ then through node $j$. We admit only routes going horizontally first then vertically.

$\mathbb{E}[W_{i,j}]$ decreases from $\frac{2}{n\sqrt{n}}$ to $o(\frac{1}{n^2})$ as a function of $\ell_{ij}$. To get a normalized expression with respect to $\sqrt{n}$, we use the coefficient $\delta_{ij}$ defined in the statement of Lemma 1.

$$\frac{2}{n\sqrt{n}}\left(1-\delta_{ij}+\delta_{ij}\ln\delta_{ij}\right) \leq \mathbb{E}[W_{i,j}] \leq$$
$$\frac{2}{n\sqrt{n}}\left(1-\delta_{ij}+\delta_{ij}\ln\delta_{ij}+\frac{1}{\sqrt{n}}-\delta_{ij}\ln\frac{\sqrt{n}+2}{\sqrt{n}+\frac{1}{\delta_{ij}}}\right).$$

This establishes the claim. In particular, if $\delta_{ij}=1/2$, then $\mathbb{E}[W_{i,j}] \sim \frac{1-\ln 2}{n\sqrt{n}}$.

■

*3) Bounding $\lambda_2(\mathbb{E}[W])$:* We need now to upperbound the second largest eigenvalue in magnitude of $\mathbb{E}[W]$, or equivalently, the relaxation time $1/(1-\lambda_2(\mathbb{E}[W]))$.

*Lemma 2 (Relaxation time):*

$$\frac{1}{1-\lambda_2(\mathbb{E}[W])} = O(\sqrt{n}). \tag{15}$$

*Proof:* The Poincaré inequality (Theorem 3) bounds the second largest eigenvalue of a stochastic matrix and not necessarily its second largest eigenvalue *in magnitude*, which is the important quantity involved in Eq. (5). It could happen that the smallest negative eigenvalue is larger in magnitude than the second largest eigenvalue. Consequently, if we show that all the eigenvalues of $\mathbb{E}[W]$ are positive, then the two eigenvalues coincide and we can use the Poincaré inequality to bound the second largest eigenvalue in magnitude. $\mathbb{E}[W]$ is symmetric so all its eigenvalues are real. The sum of all the entries along the lines of $\mathbb{E}[W]$ without counting the diagonal element is $O(1/\sqrt{n})$, whereas the diagonal elements are $\Theta(1)$, so by Gershgorin bound [4], all the eigenvalues of $\mathbb{E}[W]$ are positive.

We can now use the bounds on $\mathbb{E}[W]$ to bound its spectral gap.

We want to prove that path averaging performs $\sqrt{n}$ better than geographic gossip, where $\mathbb{E}[W_{i,j}]=1/n^2$ (IV-C.1). It is encouraging to note that for $\delta_{ij} \leqslant 1/2$, $\mathbb{E}[W_{i,j}] \geqslant \frac{1-\ln 2}{n\sqrt{n}}$, which is precisely $\sqrt{n}$ better than $1/n^2$. We thus observe that it is possible to find edges with a good capacity with length equal to half of the whole graph. However very distant destinations remain problematic. Consider the extreme case of a distance $\sqrt{n}$ between two nodes $i$ and $j$. There are only two routes that will jointly average them: the route that goes from $i$ to $j$, and the reverse one. These routes are selected with probability $1/n^2$ and $W_{ij}=1/\sqrt{n}$, implying that $\mathbb{E}[W_{ij}]=2/n^{2.5} \ll 1/n^{1.5}$.

Formally, for each ordered and distinct pair $(i,j)$, we choose a 2-hop path $\gamma_{ij}$ from $i$ to $j$ stopping by an "airport" node $k$ chosen to be located approximatively half way between $i$ and $j$. To be more precise, we define direction functions $\sigma_x$ and $\sigma_y$, where $\sigma_x(i,j)=1$ (respectively, $\sigma_y(i,j)=1$) if the horizontal (resp., vertical) part of the route from $i$ to $j$ goes to the right (resp., up) and $\sigma_x(i,j)=-1$ (resp., $\sigma_y(i,j)=-1$) if it goes left (resp., down). The coordinates of $k$ in the torus are:

$$\begin{aligned} k_x &= \left(i_x+\sigma_x(i,j)\lfloor\frac{|j_x-i_x|_{\mathcal{T}}}{2}\rfloor\right) \pmod{\sqrt{n}} \qquad (16)\\ k_y &= \left(i_y+\sigma_y(i,j)\lfloor\frac{|j_y-i_y|_{\mathcal{T}}}{2}\rfloor\right) \pmod{\sqrt{n}}. \end{aligned}$$

In the road map $\gamma$ we have just constructed, the maximum flight distance is smaller than $\frac{\sqrt{n}}{2}+1$ in $L_1$ distance. Therefore, according to Lemma 1, for any edge $e$ in $\gamma$, $\mathbb{E}[W_e] \geqslant \eta/n^{1.5}$, where $\eta$ is a non negative constant slightly smaller than $1 - \ln 2$. Thus, for each path $\gamma_{ij}$ we have:

$$\begin{aligned} |\gamma_{ij}| &= \frac{1}{\pi(i)\mathbb{E}[W_{i,k}]} + \frac{1}{\pi(k)\mathbb{E}[W_{k,j}]} \\ &= n\left(\frac{1}{\mathbb{E}[W_{i,k}]} + \frac{1}{\mathbb{E}[W_{k,j}]}\right) \\ &\leq \frac{2n^2\sqrt{n}}{\eta}. \end{aligned} \quad (17)$$

We can now compute the Poincaré coefficient:

$$\kappa = \max_e \sum_{\gamma_{ij}\ni e} |\gamma_{ij}|\pi_i\pi_j = \frac{1}{n^2}\max_e \sum_{\gamma_{ij}\ni e} |\gamma_{ij}|. \quad (18)$$

To compute this sum, we need to count the number of paths $\gamma_{ij}$ in the road map that use a given flight $e$. In our construction, we have balanced the traffic load over all the short flights so that a flight $e$ belongs to at most $8$ paths. Indeed, if a path contains flight $e$, then $e$ is either the first or second flight. In the first case, by construction, the second flight has to be approximatively as long as $e$. Moreover, because of quantized grid effects, there are actually only $4$ different possible flights a traveler in flight $e$ might take as second flight (see Fig. 8). Repeating this argument in the case where $e$ is the second flight, we then obtain that a flight $e$ appears in at most $8$ paths. Combining (17) and (18), we get:

$$\kappa \leq \frac{16}{\eta}\sqrt{n}.$$

As a result,

$$\lambda_2 \leq 1 - \frac{\eta}{16\sqrt{n}},$$

which yields Lemma 2. The proof is complete by using equation (5).

■

In the next Section, we generalize this proof from grids to regular geometric graphs. The approach will be the same but the detailed computations will be different. Also, the construction of the paths in the travel agency method will need some refinement.

### *B. Performance of box-path averaging.*

We now prove Theorem 5. All the fundamental ideas coming from the proof on grids in the previous section, appear here again, but sometimes in a more technical form. We have $k$ boxes forming a torus grid as in the previous section and $k = \lceil\sqrt{(n/(\alpha\log n))}\rceil^2 \simeq n/(\alpha\log n)$, for some $\alpha > 2$.

Using regularity, each box contains a number of nodes between $a\log n$ and $b\log n$. We use the $(\leftrightarrow, \updownarrow)$-box routing scheme presented in Section III-D. There are only a few modifications to make to the grid proof in order to obtain the regular geometric graph proof. The idea is to notice that for any route $r = (r_1, r_2, \cdots, r_\ell)$, we can attribute a box route $\widetilde{r}$ consisting of the boxes the nodes of $r$ belong to. If we call $b(i)$ the box node $i$ belongs to, then $\widetilde{r} = (b(r_1), b(r_2), \cdots, b(r_\ell))$. We call $n_i$ the number of nodes in the box $b(i)$ node $i$ belongs to. The sequence of $n_i$ is fixed by the graph we are considering. $\ell_{ij}$ is the $L_1$ distance between boxes $b(i)$ and $b(j)$: $\ell_{ij} = \|b(j) - b(i)\|_1$.

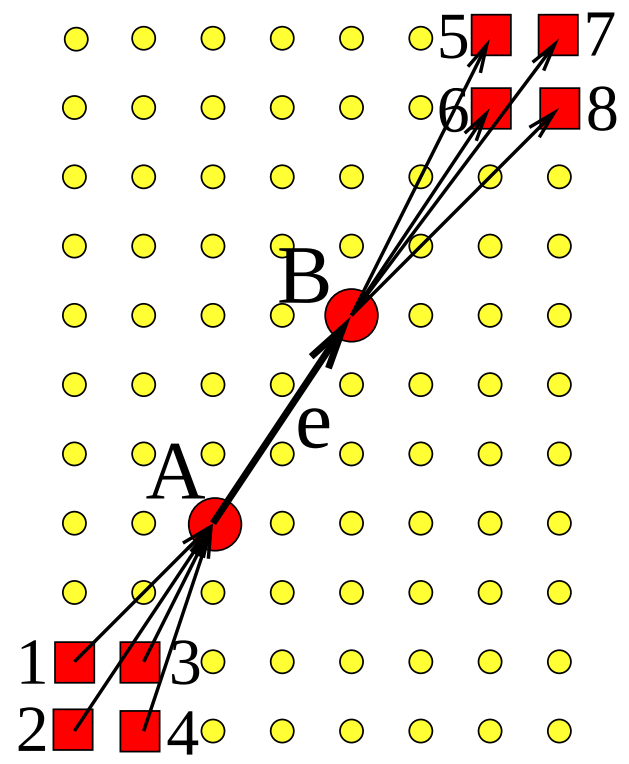

Fig. 8. Number of paths including an edge $e = (A, B)$ in the road map. Paths have two hops of equal length, where equality here is defined up to grid effects. Therefore, for a given edge $e$, there are at most 8 paths including $e$: $(1, A, B)$, $(2, A, B)$, $(3, A, B)$, $(4, A, B)$ and $(A, B, 5)$, $(A, B, 6)$, $(A, B, 7)$, $(A, B, 8)$.

We denote by $\ell(r)$ the number of nodes in route $r$, $s(\widetilde{r})$ the starting box of route $\widetilde{r}$ and $d(\widetilde{r})$ its destination box. In our problem the chosen route is random, which we will denote by capital case letter: $R$, leading to other random variables $\widetilde{R}$, $\ell(R)$, $s(\widetilde{R})$, etc.

*1) Evaluating* $\mathbb{E}[W]$*:*

*Lemma 3: (Expected* $\mathbb{E}[W]$ *on the regular geometric graph)* For any pair of nodes $(i, j)$ that do not belong to the same box, if their grid-distance normalized to the maximum grid-distance $\delta_{ij} = \ell_{ij}/\sqrt{k}$ is smaller than a constant, then

$$\mathbb{E}[W_{ij}] = \Omega\left(\frac{1}{n\sqrt{n\log n}}\right). \tag{19}$$

More precisely,

$$\mathbb{E}[W_{i,j}] \geq \frac{4a}{b^2}\frac{2}{n^2}\sqrt{\frac{n}{\alpha\log n}}\left(1-\delta_{ij}+\delta_{ij}\log\delta_{ij}\right), \tag{20}$$

*Proof:* For any node $i$ and node $j$ that do not belong to the same box, we want to compute the expectation of $W_{ij}$. Counting the routes in this setting is complicated because each sender has at least $a\log n$ nodes to send its message to. In order to use our simple analysis of the grid, we condition the expectation on the box routes $\widetilde{R}$. Given a box route, $W_{ij} = 0$ if $i$ or $j$ is not in the box route. On the contrary, if they both are in the box route, then $W_{ij} = 1/\ell(\widetilde{R})$ with probability $1/(n_i n_j)$. Indeed, if $i$ (or $j$) is in starting box, the probability that $i$ is the starting node is $1/n(i)$, because all the nodes wake up with the same rate. If $i$ (or $j$) is in another box of the given box route, then the probability that $i$ is chosen is $1/n(i)$ as well, because the routing chooses next node uniformly among the nodes of the next box.

$$\begin{aligned}\mathbb{E}[W_{ij}] &= \mathbb{E}_{\widetilde{R}}[\mathbb{E}_R[W_{ij}|\widetilde{R}]]\\ &= \mathbb{E}_{\widetilde{R}}[\frac{1}{n_i n_j}\frac{1}{\ell(\widetilde{R})}1_{b(i)\in\widetilde{R}}1_{b(j)\in\widetilde{R}}].\end{aligned}$$

From now on, we are back to a problem with routes on a grid which has $k$ "nodes". The difference with previous section is that routes are no longer uniform. Indeed, now, boxes wake up more frequently if they contain more nodes: the probability that box $b_i$ wakes up is $n_i/n$. Destination boxes are still chosen uniformly at random with probability $1/k$ because there are $k$ boxes in total. Just as before, we consider only ($\leftrightarrow, \updownarrow$)-box routes so that a box route is entirely determined by its starting box and its destination box, and we count box routes of different length separately. Let $\mathcal{R}_{ij}^{\ell}$ be the set of box routes of size $\ell$ including $b_i$ and $b_j$.

$$\begin{aligned}
\mathbb{E}[W_{ij}] &= \frac{1}{n_i n_j}\sum_{\widetilde{r}} \frac{1_{b(i)\in\widetilde{r}}1_{b(j)\in\widetilde{r}}}{\ell(\widetilde{r})}\mathbb{P}[\widetilde{R}=\widetilde{r}] \\
&= \frac{1}{n_i n_j}\sum_{\ell=\ell_{ij}+1}^{2\lfloor\frac{\sqrt{k}}{2}\rfloor+1}\sum_{\widetilde{r}\in\mathcal{R}_{ij}^{\ell}} \frac{\mathbb{P}[\widetilde{R}=\widetilde{r}]}{\ell} \\
&= \frac{1}{n_i n_j}\sum_{\ell=\ell_{ij}+1}^{2\lfloor\frac{\sqrt{k}}{2}\rfloor+1}\sum_{\widetilde{r}\in\mathcal{R}_{ij}^{\ell}} \frac{\mathbb{P}[s(\widetilde{R})=s(\widetilde{r}),d(\widetilde{R})=d(\widetilde{r})]}{\ell} \\
&= \frac{1}{n_i n_j}\sum_{\ell=\ell_{ij}+1}^{2\lfloor\frac{\sqrt{k}}{2}\rfloor+1}\sum_{\widetilde{r}\in\mathcal{R}_{ij}^{\ell}} \frac{1}{\ell}\frac{n_{s(\widetilde{r})}}{n}\frac{1}{k}.
\end{aligned}$$

We now use the regularity of the graph : for any node $m$, $a\log n \leqslant n_m \leqslant b\log n$.

$$\begin{aligned}
\mathbb{E}[W_{ij}] &\geqslant \frac{1}{(b\log n)^2}\sum_{\ell=\ell_{ij}+1}^{2\lfloor\frac{\sqrt{k}}{2}\rfloor+1}\frac{1}{\ell}\frac{a\log n}{n}\frac{4\log n}{n}|\mathcal{R}_{ij}^{\ell}|. \\
&= \frac{4a}{b^2}\frac{1}{n^2}\sum_{\ell=\ell_{ij}+1}^{2\lfloor\frac{\sqrt{k}}{2}\rfloor+1}\frac{|\mathcal{R}_{ij}^{\ell}|}{\ell} \\
&\geqslant \frac{4a}{b^2}\frac{2}{n^2}\left(\sqrt{k}-\ell_{ij}-\ell_{ij}\ln\frac{\sqrt{k}}{\ell_{ij}}\right).
\end{aligned}$$

The last inequality comes from the same computation as for the grid, and it can be reformulated as in Lemma 3 when using the normalized distance coefficient $\delta_{ij}=\ell_{ij}/\sqrt{k}$. ■

*2) Bounding $\lambda_2(\mathbb{E}[W])$:*

*Lemma 4 (Relaxation time RGG):*

$$\frac{1}{1-\lambda_2(\mathbb{E}[W])} = O(\sqrt{n\log n}). \tag{21}$$

*Proof:* As for the grid, we now apply the travel agency method. The situation is very similar to the grid case, except that boxes now contain $\Theta(\log n)$ nodes each.

Similarly to the grid case, we will be using 2-hop paths for every pair of nodes, by adding one intermediate stop half-way. More precisely, this intermediate stop is chosen in the box whose coordinates on the underlying lattice are given by equations 16, where $i$ and $j$ are the lattice coordinates of the source and destination boxes. Then, within each box, we need to carefully and fairly assign the intermediate nodes because a flight should not be used more than a constant number of times (it was 8 for the grid), otherwise it would create congestion. It is not hard to design such road maps because the number of nodes in each box varies at most by a constant multiplicative factor $b/a$.

To show this, assume that each box contains exactly $\log n$ nodes. Then, there are $(\log n)^2$ road maps to find between all the nodes in a pair of boxes (assume box $1$ and $3$, and let box $2$ be the one half-way), but happily enough, there are $(\log n)^2$ flights between box $1$ and box $2$ and also between box $2$ and $3$. Therefore, as we can see on Fig. 9, the box path (box $1$, box $2$, box $3$) can correspond to $(\log n)^2$ node road maps all using different flights (edges). This flight allocation technique can easily be extended to cases where the boxes do not have the same number of airports by using some flights at most $\lceil b/a \rceil$ times each in the paths between two given boxes.

There is a second refinement to the grid proof: solving the problem for nodes that share a common box, which do not average jointly (Our bound on $\mathbb{E}[W_{ij}]$ is zero). However there are many edges to nodes in neighboring boxes. So formally, if node $i$ and node $j$ are in the same box, we design the road map from $i$ to $j$ to be a two hop road map stopping at a node located in the box above their box. By sharing fairly the available relay airports, the short north-south flights might be used in $\lceil b/a \rceil$ extra road maps.

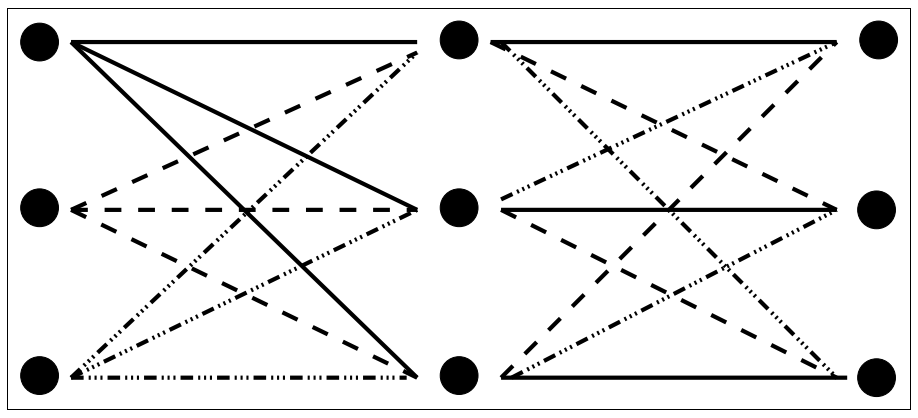

Fig. 9. Path allocation when there are 3 nodes per box and thus 9 paths to design.

We can thus construct road maps for any pair of airports that will use at most $9\lceil b/a \rceil$ times each good intensity flight. The rest of the proof is identical to the grid proof.

For each path we have:

$$\begin{aligned} |\gamma_{ij}| &= \frac{1}{\pi(i)\mathbb{E}[W_{i,k}]} + \frac{1}{\pi(k)\mathbb{E}[W_{k,j}]} \\ &= n\left(\frac{1}{\mathbb{E}[W_{i,k}]} + \frac{1}{\mathbb{E}[W_{k,j}]}\right) \\ &\leq cn^2\sqrt{n\log n}, \end{aligned} \tag{22}$$

for some constant $c$. Inequality 22 was obtained with the same reasoning as in the grid. We therefore conclude, using the Poincaré coefficient argument that

$$\kappa \leq 9\lceil \frac{b}{a} \rceil c\sqrt{n\log n}.$$

As a result, for $n$ large enough, and some constant $c'$.

$$\lambda_2 \leq 1 - \frac{1}{c'\sqrt{n\log n}},$$

which yields the lemma. ■

### C. Regularity of random geometric graphs

*Lemma 5 (Regularity of random geometric graphs):* Consider a random geometric graph with $n$ nodes and partition the unit square in boxes of size $\alpha\frac{\log n}{n}$. Then, all the boxes contain $\Theta(\log n)$ nodes, with high probability as $n \to \infty$.

*Proof:* Let $X_i$ denote the number of nodes contained in the $i$th box. $X_i$ are (non-independent) Binomially distributed random variables with expectation $\alpha \log n$. Standard Chernoff (we do not optimize for the constants) bounds [16] imply:

$$\mathbb{P}(X_i \leq \frac{\alpha}{2}\log n) \leq e^{-\alpha/8 \log n}.$$

and

$$\mathbb{P}(X_i \geq 2\alpha \log n) \leq e^{-\alpha/3 \log n}.$$

which give tight bounds on the number of nodes in each box:

$$\mathbb{P}(\frac{\alpha}{2}\log n \leq X_i \leq 2\alpha \log n) \geq 1 - 2e^{-\alpha/8 \log n}. \tag{23}$$

A union bound over boxes yields the uniform bounds on the maximum and minimum load of a square:

$$\mathbb{P}(\frac{\alpha}{2}\log n \leq \min_i X_i \leq \max_i X_i \leq 2\alpha \log n) \geq 1 - n^{1-\alpha/8}\frac{2}{\alpha \log n}.$$

Therefore, selecting $\alpha \geq 8$ yields the lemma. A more technical proof shows that the lemma holds for $\alpha > 2$. ∎